\documentclass[10pt,a4paper]{article}

\usepackage[english]{babel}
\usepackage[T1]{fontenc}
\usepackage[dvips]{graphics}
\usepackage{epsfig}
\usepackage{latexsym}
\usepackage{array}
\usepackage{color}

\usepackage{amsmath}
\usepackage{amssymb}

\usepackage{amsfonts}
\usepackage{euscript}

\title{Magnetic shielding properties of high-temperature superconducting tubes subjected to axial fields}
\author{S~Denis$^{1,2}$, L~Dusoulier$^{1,3}$, M~Dirickx$^{1}$, Ph~Vanderbemden$^{2}$, R~Cloots$^{3}$, M~Ausloos$^{4}$ \\ and B~Vanderheyden$^{2}$}

\begin{document}
\maketitle

\section{abstract}
We have experimentally studied the magnetic shielding properties of a cylindrical shell of BiPbSrCaCuO subjected to low frequency AC axial magnetic fields. The magnetic response has been investigated as a function of the dimensions of the tube, the magnitude of the applied field, and the frequency. These results are explained quantitatively by employing the method of E. H. Brandt (Brandt E H 1998 {\it Phys. Rev. B} {\bf 58} 6506) with a $J_c(B$) law appropriate for a polycrystalline material. 
Specifically, we observe that the applied field can sweep into the central region either through the thickness of the shield or through the opening ends, the latter mechanism being suppressed for long tubes.
For the first time, we systematically detail the spatial variation of the shielding factor (the ratio of the applied field over the internal magnetic field) along the axis of a high-temperature superconducting tube. 
The shielding factor is shown to be constant in a region around the centre of the tube, and to decrease
as an exponential in the vicinity of the ends. This spatial dependence comes from the competition between  two mechanisms of field penetration. 
The frequency dependence of the shielding factor is also discussed and shown to follow a power law arising from the finite creep exponent $n$.

%--------------------------------------------------------------------
% !!!!!!!!! noter partout d'où viennent les figures !! de quelles sim ça provient
\section{Introduction}
%--------------------------------------------------------------------
Electromagnetic shielding has two main purposes. The first one is to
prevent an electronic device from radiating electromagnetic energy, in
order to comply with radiation regulations, to protect neighbouring
equipments from electromagnetic noise, or, in certain military
applications, to reduce the electromagnetic signature of the
device. The second purpose of shielding is to protect sensitive
sensors from radiation emitted in their surroundings, in order to take
advantage of their full capabilities.

As long as the frequency of the source field remains large, typically
$f > 1~\mathrm{kHz}$, conducting materials can be used to attenuate
the field with the skin effect. For the lowest frequencies, however,
 conductors continue to act as good electric shields (and can
be used to make a Faraday cage), but they fail to shield magnetic
fields. The traditional approach to shield low frequency magnetic
fields consists in using soft ferromagnetic materials with a
high relative permeability, which divert the source field
from the region to protect~\cite{3}. As the magnetic permeability decreases
with increasing frequency, this approach is only practical for low
frequencies (typically $f < 1~\mathrm{kHz}$). If low temperatures are
allowed by the application (77 K for cooling with liquid nitrogen),
shielding systems based on high-temperature superconductors (HTS)
compete with the traditional solutions~\cite{revue Pavese}. Below
their critical temperature, $T_c$, HTS are strongly diamagnetic and
expel a magnetic flux from their bulk. They can be used to construct
enclosures that act as very effective magnetic shields over a broad
frequency range~\cite{revue Pavese}.

Several factors determine the quality of a HTS magnetic shield. First,
a threshold induction, $B_\mathrm{lim}$, characterizes the maximum
applied induction that can be strongly attenuated. In the
case of a shield that is initially not magnetized and is subjected to
an increasing applied field, the internal field remains close to zero until the
applied induction rises above $B_\mathrm{lim}$. The field then
penetrates the inner region of the shield and the induction increases
with the applied field~\cite{160, 110, 108, 117, 109, 113}. 
A second important factor is the geometrical
volume over which a shield of given size and shape can attenuate an
external field below a given level. A third determining factor is the
frequency response of the shield.

In this paper, we focus on the shielding properties of a ceramic
tube in the parallel geometry, which means that the source field is
applied parallel to the tube axis. This geometry is amenable to direct
physical interpretation and numerical simulations, as currents flow
along concentric circles perpendicular to the axis. Note that a
HTS tube certainly outperforms a ferromagnetic shield in the parallel
geometry~\cite{70}. For a ferromagnetic tube with an infinite length, the shield
does not attenuate the external magnetic field since its longitudinal
component must be continuous accross the air-ferromagnet
interface. For finite lengths, the magnetic flux is caught in the
material because of demagnetization effects but the shielding
efficiency remains poor for long tubes.

A number of results can be found in the literature on HTS tubes in
the parallel geometry. For HTS polycrystalline materials,
$B_{\mathrm{lim}}$ was found to vary between $0.3~\mathrm{mT}$ for a
tube with a superconducting wall of thickness $d =
40~\mu\mathrm{m}$~\cite{170,moieucas}, and $15~\mathrm{mT}$ for $d =
2.2~\mathrm{mm}$~\cite{113} at $77~\mathrm{K}$. 
If lower temperatures are allowed than 
$77~\mathrm{K}$, 
higher $B_{\mathrm{lim}}$ values can be obtained with other compounds.
As an example, MgB$_2$ tubes were reported to shield magnetic inductions up to $1~\mathrm{T}$ at $4.2~\mathrm{K}$~\cite{198,240}.
Results on the variation of the field
attenuation along the axis appear to be contradictory. An exponential
dependence was measured for a YBCO tube~\cite{173} and for a BSCCO tube ~\cite{112}.
Other measurements~\cite{109,196} in similar
conditions have shown instead a constant shielding factor in a region
around the centre of YBCO and BSCCO tubes. 
As for the frequency response, the
shielding factor is expected to be constant if flux creep effects are
negligible, as is the case in Bean's model~\cite{Bean1,Bean2}. It is on
the other hand expected to increase with frequency in the presence of
flux creep, since the induced currents saturate to values that
increase with frequency~\cite{144}. Experimental data have shown very
diverse behaviours.  
In \cite{112}, the field attenuation due to a
thick BSCCO film on a cylindrical silver substrate was found to be
frequency independent. The same results were established for superconducting disks
made from YBCO powder and subjected to perpendicular
fields~\cite{119,45}. Yet other studies on bulk BSCCO
tubes~\cite{160,218} measured a field attenuation that
decreases with frequency, whereas the attenuation was shown to slowly
increase with frequency for a YBCO superconducting
tube~\cite{150}.

The purpose of this paper is to provide a detailed study of 
the magnetic shielding properties of a polycrystalline HTS tube, with regard to
the three determining factors: threshold induction, spatial
variation of the field attenuation, and frequency response. The study is
carried both experimentally and by means of numerical simulations, in
order to shed light on the relation between the microscopic mechanisms
of flux penetration and the macroscopic properties. For the numerical
simulations, we have followed the method proposed by E.~H.~Brandt in
\cite{144}, which can be carried easily with good precision on a
personal computer. We focus on a HTS tube with one opening at each
end and assume that the superconducting properties are uniform along
the axis and isotropic.

The report is organized as follows.  The sample and the
experimental setup are described in section \ref{samples}. In section \ref{compute}, we
discuss the constitutive laws that are appropriate for a polygrain
HTS, set up the main equations and the numerical model.  Section \ref{s:DC-mode} is
devoted to the shielding properties of superconducting tubes subjected
to slowly time varying applied fields (called the DC mode). First, the
evolution of the measured internal magnetic induction of a commercial
sample versus the applied induction is presented. We then detail the
field penetration into a HTS tube 
and study the field attenuation as a function of
position along the tube axis. The frequency response of the shield is addressed in section \ref{s:AC-mode},
where it is shown that the variations with frequency can be explained by
scaling laws provided heat dissipation can be neglected. Our main
results are summarized in section \ref{conclusions}, where we also draw conclusions of
practical interest.

%parler de Blim; dire qu'il n'est pas bien défini. Nous on discute ce Blim en fction des caractéristiques géo et on le relie à Hp. On explique égalemt physiqmt quels sont les paramètres géométriques qui peuvent influencer ce Blim : effets démagnétisants, épaisseur paroi. Cela est discuté grâce aux lignes de H et fronts Jc=0.5

%il y a différentes études qui ont étudié la variation de SF le long de l'axe d'un tube supra. Cependant, contradictions. Ici, on explique clairement ce qu'il se passe grâce aux simulations. On montre égalmt que l'on retrouve ce cptmt lors de la mesure en AC

%le cptmt fréquentiel est important d'un point de vue pratique. Certaines études rapportent des SF qui diminuent qd f augmente. Sûremt dû à échauffemt. Nous, on montre que s'il n'y a pas d'échauffement, SF(f) suit une loi en puissance. On retrouve cette loi en appliquant les lois d'échelles provenant de J^n. 

% dire qu'ici on montre que l'algo de Brandt permet de décrire cptmts expérimentaux et que cette méthode est élégante et facile à mettre en oeuvre => éventuelmt, parler d'autres méthodes qui me semblent moins bonnes ... (il y a un article de IEEE discutant de l'uniformité du chp à l'intérieur d'un tube supra : article mauvais).

%\hrule
%\vspace{0.3cm}
%\begin{center}
%SAMUEL : mes commentaires et questions sont en ``{\sc small caps}''.
%\end{center}
%\vspace{0.3cm}
%\hrule
%\vspace{0.3cm}
%\vspace{0.3cm}

%----------------------------------------------------
\section{Experimental}\label{samples}
%----------------------------------------------------

  \begin{table}
    \caption{\label{t:sample-description} 
      Physical characteristics of the sample: the material composition and the critical temperature come from \cite{CAN}}
%   \begin{indented}
%\item[]
   \begin{tabular}{@{}ll}
      \hline
      Material & Bi$_{1.8}$Pb$_{0.26}$Sr$_2$Ca$_2$Cu$_3$O$_{10+x}$ \\
      Length & $\ell  =8~\mathrm{cm}$ \\
      Inner radius & $a_1=6.5~\mathrm{mm}$\\
      Outer radius & $a_2=8~\mathrm{mm}$\\
      Wall thickness & $d= 1.5~\mathrm{mm}$\\
      Critical temperature & $T_c \cong 108~\mathrm{K}$\\
     \hline
    \end{tabular}
  % \end{indented}
\end{table}
We measured the shielding properties of a commercial superconducting
specimen (type CST-12/80 from CAN Superconductors), which was
cooled at $T=77~\mathrm{K}$ under zero-field. The sample is a tube
made by isostatic pressing of a polygrain ceramic. 
Its main characteristics are summarized in
table~\ref{t:sample-description}.

\begin{figure}[ht]
\center \includegraphics[scale=0.5]{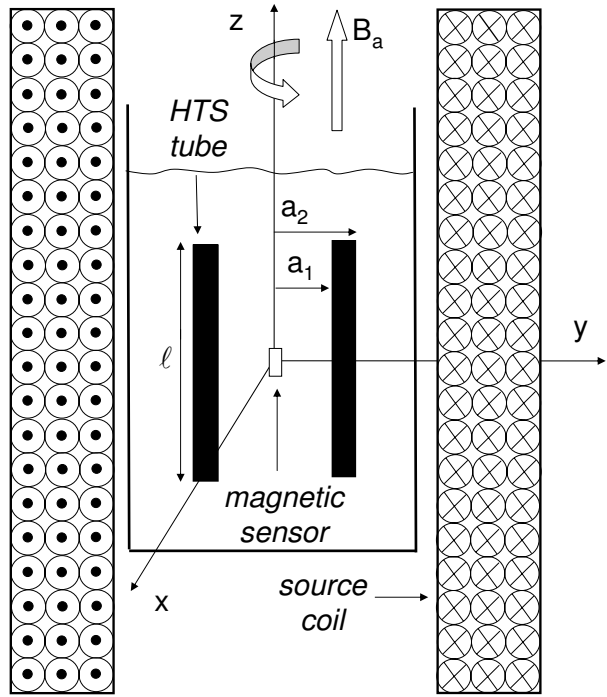}\caption {Experimental
setup. The superconducting tube of length $\ell$ and wall tickness
$d=a_2-a_1$ is placed inside a coil generating an axial induction of
magnitude $B_a$. The applied field can have the form of a slow ramp,
in which case the magnetic sensor is a Hall probe connected to a
nanovoltmetre, or a that of a low-frequency alternating field, in which
case the field inside the tube is measured by a pick-up coil connected
to a lock-in amplifier.  In all cases, the sample and the sensor are
cooled with liquid nitrogen (T = 77~K) under zero-field
condition.}\label{figure1}
\end{figure}

The experimental setup is shown in figure~\ref{figure1}. The sample is
immersed in liquid nitrogen and placed inside a source coil generating
an axial magnetic induction $\mathbf{B}_a = B_a\,\hat{z}$.  The
applied induction, $B_a$, can be generated in two different modes. In the first
mode, called the DC mode, $B_a$ increases at a constant rate of 
$\dot B_a\cong 0.2~\mathrm{mT/s}$ with a brief stop (around 1 second) needed to measure the internal induction at each wanted value of $B_a$.
The maximum applied induction in this mode is $30~\mathrm{mT}$.
The induction in the inside of the shield, $B_{\mathrm{in}}$, is measured with a
Hall probe placed in the centre of the tube; the probe is connected to a HP34420 nanovoltmetre. 
To reduce noise from outside sources, the setup is enclosed in a double mu-metal ferromagnetic shield.
The field resolution is around $1~\mathrm{\mu T}$.
In the second mode of operation, called the AC mode, the applied field is a low-frequency
alternating field with no DC component. The frequency of the applied induction ranges between $43~\mathrm{Hz}$ and $403~\mathrm{Hz}$ and the amplitude of $B_a$ can reach $25~\mathrm{mT}$.
The field inside the tube is
measured by a pick-up coil, which can be moved along the $z-$axis and
whose induced voltage is measured with an EGG7260 lock-in amplifier. In this
mode, the setup can measure magnetic inductions as weak as $1~\mathrm{nT}$ at $103~\mathrm{Hz}$. As a
result, care must be taken to reject common-mode electrical
noise. In the present work, the capacitive coupling between the source and the
pick-up coils was reduced by electrically connecting the
superconducting tube to ground so as to realize an electrical shield.
%----------------------------------------------------
\section{Theory}\label{compute}
%----------------------------------------------------

\subsection{Flux penetration in polycrystalline bulk ceramics}

Bulk polycrystalline BiSrCaCuO ceramics consist of a stack of a large number of superconducting grains~\cite{224}. The penetration of a magnetic flux in
such a material is inhomogeneous and strongly depends on the
microstructure, as shielding currents can flow both in the grains and the
intergranular matrix~\cite{225}. For a polygrain material that has
been cooled in zero-field condition, the flux penetrates in roughly
three different steps~\cite{169}. First, for the weakest applied
fields, Meissner surface currents shield the volume and no flux enters
the sample.  When the local induction, $B$, exceeds
$\mu_0\,H_{c1j}$, where $H_{c1j}$ is the lower critical field of the
intergranular matrix, vortices start
entering this region. The magnetic flux penetrates the grains 
at the higher induction $B\sim\mu_0 H_{c1g}$~\cite{194},
where $H_{c1g}$ is the lower critical field of the grains themselves.

\subsection{Model assumptions}

In our model, we neglect surface barrier effects and set $H_{c1j}$ to
zero. Therefore, flux starts threading the intergranular matrix as
soon as the applied field is turned on. The penetration of individual
grains depends on the intensity of the local magnetic field, which,
because of demagnetization effects, varies as a function of the grain
sizes and orientations. The penetration of each grain may thus
take place over a range of applied fields: we expect an
increasing number of grains to be penetrated as the external field is
increased. Since we aim at studying the macroscopic properties of the
superconducting tube and aim at deriving recommandations of practical interest,
we will not seek to describe grains individually and thus neglect
detailed effects of their diamagnetism. We will instead consider the
induction $\mathbf{B}$ to be an average of the magnetic flux over many
grains and assume the constitutive law
$\mathbf{B}=\mu_0\,\mathbf{H}$. The resulting model describes the
magnetic properties of an isotropic material which supports
macroscopic shielding currents.

We will further assume the material to obey the conventional~\cite{144,148} constitutive
law
\begin{equation}\label{Jexpn}
\mathbf{E}(J)=E_{c}\left({J \over J_c}\right)^n\:\frac{\mathbf{J}}{J},
\end{equation}
where $J$ is the module of the vector current density $\mathbf{J}$.
The exponent $n$ allows for flux creep and typically ranges from
10 to 40 for YBCO and BSCCO compounds at 77~K. The value
for $n$ that is adequate for the sample of
table~\ref{t:sample-description} is to be determined from the
frequency dependence of its shielding properties, see
section~\ref{ss:AC-mode-simulations}.  Note that one recovers Bean's
model, which neglects flux creep effects, by taking the limit
$n\rightarrow\infty$. A final constitutive law comes from the
polygrain nature of the material. The critical current density is
assumed to decrease with the local induction as in Kim's
model~\cite{211}:
\begin{equation}\label{Kim}
J_c(B)=\frac{J_{c0}}{1+B/B_1},
\end{equation} 
where $J_{c0}$ and $B_1$ are experimentally determined by fitting
magnetization data, as discussed in section~\ref{ss:DC-discussion}.

\subsection{Model equations and numerical algorithm}
\label{ss:algo}

A common difficulty in modelling the flux penetration in HTS materials
arises from the fact that the direction of the shielding currents is
usually not known {\it a priori} and, furthermore, may evolve over
time as the flux front moves into the sample. This problem is greatly
simplified for geometries in which the direction of the shielding
currents is imposed by symmetry. Examples include long bars in a
perpendicular applied field \cite{148}, in which case the currents flow
along the bar, and axial symmetric specimens subjected to an axial
field, for which the currents flow along concentric circles
perpendicular to the symmetry axis. Numerous examples have been
extensively studied by E.~H.~Brandt~\cite{171} for both geometries, by means of a
numerical method based on the discretization of Biot-Savart integral
equations. In this work, we follow Brandt's method
for modelling the flux penetration in a tube subjected to an axial
field.

To set up the main equations, we closely follow \cite{144}. As a
reminder, the sample is a tube of internal radius $a_1$, external
radius $a_2$, and length $\ell$ (see figure \ref{figure1}). We work with
cylindrical coordinates, so that positions are denoted by
$(r,\varphi,z)$. As the magnitude of the axial induction, $B_a$, is
increased, the induced electric field and the resulting current
density assume the form
\begin{equation}\label{J et E}
\mathbf{J}=-J(r,z)\,\hat{\varphi},\quad \mathbf{E}=-E(r,z)\,\hat{\varphi}, 
\end{equation}
where $\hat{\varphi}$ is the unit vector in the azimuthal
direction. The magnetic induction is invariant under a rotation around
the $z$-axis and has no $\varphi$-component. Thus,
\begin{equation}\label{B}
\mathbf{B}(r,z)=B_r(r,z)\,\hat{r}+B_z(r,z)\,\hat{z}. 
\end{equation}
The fields $\mathbf{B}$, $\mathbf{E}$, and the current density,
$\mathbf{J}$, satisfy Maxwell's equations
\begin{eqnarray}
  \nabla\times\mathbf{E} & = & - \frac{\partial \mathbf{B}}{\partial t} ,\label{Maxwell-rotE}\\
  \nabla\times\mathbf{B} & = & \mu_0\,\mathbf{J},\label{Maxwell-rotB}
\end{eqnarray}
where we have used the constitutive law $\mathbf{B} =
\mu_{0}\mathbf{H}$.

In order to avoid an explicit and costly computation of the magnetic
induction $\mathbf{B}(\mathbf{r},t)$ in the infinite region exterior
to the tube, an equation of motion is first established for the
macroscopic shielding current density $\mathbf{J}(\mathbf{r},t)$,
since its support is limited to the volume of the
superconductor.  The magnetic field is then obtained where required by
integrating the Biot-Savart law. After having eliminated $\mathbf{B}$
and integrated over $\varphi$, this procedure leads to the integral
equation~\cite{144}
\begin{equation}\label{24}
E(\mathbf{r})=-\mu_0\int^{a_2}_{a_1}\int^{\ell/2}_{0}Q(\mathbf{r},\mathbf{r'})\dot{J}(\mathbf{r'})\mathrm{d}r'\mathrm{d}z'
+\frac{r}{2}\dot{B}_a,
\end{equation}
where $\mathbf{r}$ and $\mathbf{r'}$ are shorthands for $(r,z)$ and
$(r',z')$, while $Q(\mathbf{r},\mathbf{r'})$ is a kernel which only
depends on the sample geometry. In the present case, $Q$ assumes the
form
\begin{equation}\label{Q}
Q(\mathbf{r},\mathbf{r'})=f(r,r',z-z')+f(r,r',z+z'),
\end{equation}
where
\begin{equation}\label{f}
f(r,r',\eta)=\int^{\pi}_{0}\frac{r'\cos\varphi}
{2\pi\sqrt{\eta^2+r^2+r'^2-2rr'\cos\varphi}}\:\mathrm{d}\varphi,
\end{equation}
is to be evaluated numerically as suggested in \cite{144}. By contrast
to \cite{144}, the kernel is integrated in the radial direction
from $r^{'}=a_1$ to $r^{'}=a_2$, as dictated by the tubular geometry
of the sample.

The equation of motion for $\mathbf{J}$ is obtained in three
steps. First, the electric field is eliminated from (\ref{24}) by
using the constitutive law (\ref{Jexpn}).  Second, the equation is
discretized on a two-dimensional grid with spatial steps $\Delta r$
and $\Delta z$.  Third, the resulting matrix equation is inverted,
yielding the relation
\begin{equation}\label{34}
\dot{J}_{i}(t)=\frac{1}{\mu_0 \Delta r \Delta z}\sum_j Q_{ij}^{-1} \left\{\frac{r_j}{2}\dot{B}_a-E\left[J_{j}(t)\right]
\right\}.
\end{equation}
Here, $J_i$ and $Q_{ij}$ are shorthands for $J(\mathbf{r}_i)$ and
$Q(\mathbf{r}_i,\mathbf{r}_j)$. Actually, the two-dimensional space matrix
is transformed into a one-dimensional vector.
Imposing
the initial condition
\begin{equation}\label{initJ}
J_i(t=0)=0 \quad \forall i,
\end{equation} 
the current density can be numerically integrated over time by
updating the relation
\begin{equation}\label{J_t+dt}
J_i(t+\Delta t)\cong J_i(t)+\dot{J}_i(t)\,\Delta t,
\end{equation}
where $\dot{J}_i$ is evaluated as in (\ref{34}) and $\Delta t$ is
chosen suitably small. An adaptative time step procedure described in
\cite{144} makes the algorithm converge towards a solution that
reproduces the experimental data fairly well, see sections
\ref{s:DC-mode} and \ref{s:AC-mode}. Note that for those geometries
that have one dimension much larger than the others, as is the case for a long tube with a thin wall, one can improve
the convergence while preserving the precision 
by working with rectangular cells with the refinement described in \cite{178}.

According to the two different modes of operation of the external
source that were introduced in section~\ref{samples}, we have run the
algorithm with $B_a(t)$ either in the form of a ramp, $B_a(t) =
\dot{B}_a\,t$, or as a sinusoidal source of frequency $f$, $B_a(t) =
B_o \,\mathrm{sin}(2\pi f t)$. The shielding properties of the sample
are evaluated in both cases by probing the magnetic flux density at
points located along the $z$-axis of the tube. By
symmetry, this field is directed along $\hat{z}$, and we define $B_{z\mathrm{in}}(z)$ as
\begin{eqnarray}
  \mathbf{B}(r=0,z)=B_{z\mathrm{in}}(z)\,\hat{z}.
\end{eqnarray}
For the DC  mode, with $B_a = \dot{B}_a\,t$, we define the DC shielding factor as
\begin{equation}\label{SFDC}
DCSF(z)=\frac{B_a}{B_{z\mathrm{in}}(z)}.
\end{equation}
However, in the AC mode, with $B_a(t)=B_o \,\mathrm{sin}(2\pi f
t)$ , one must pay attention to the non-linearity of the magnetic
response of the sample. The induction $B_{z\mathrm{in}}$ contains several
harmonics, all odd in the absence of a DC component \cite{193}. We are
led to define the AC shielding factor as
\begin{equation}\label{SFAC}
ACSF(z)=\frac{B_{a,RMS}}{B_{z\mathrm{in},RMS}(z)},
\end{equation}
where $B_{a,RMS} = B_o/\sqrt{2}$ is the RMS value of the applied
magnetic induction and $B_{z\mathrm{in},RMS}(z)$ is the RMS value of the
fundamental component of $B_{z\mathrm{in}}$, which can be directly
measured by the lock-in amplifier.

%cahier labo VI p 109.
The algorithm presented in this section allows us the determine $DCSF$
up to $10^7$ in the DC mode and $ACSF$ up to $10^4$ in the
AC mode.

%----------------------------------------------------
\section{Magnetic shielding in the DC mode}                                     \label{s:DC-mode}
%----------------------------------------------------

%----------------------------------------------------
\subsection{Experimental results}
\label{ss:DC-experimental}

%144e/tube11/sim&mesuregood.eps
\begin{figure}[ht!]
\center \includegraphics[scale=0.5]{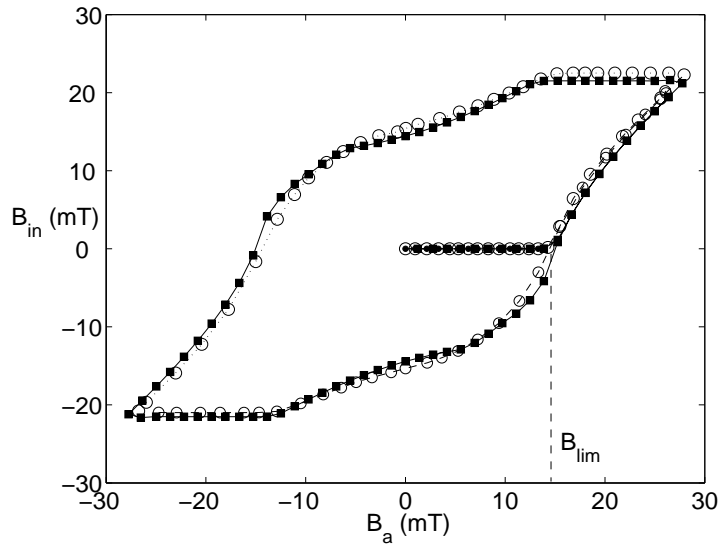}\caption {Evolution of the internal magnetic induction at the centre of the
tube, $B_{z\mathrm{in}}(z=0)$, as a function of the applied
induction. The sample (table~\ref{t:sample-description}) is cooled in zero-field conditions down to
$T=77~\mathrm{K}$. The open circles represent the experimental data
and the filled squares represent the simulation, as
discussed in subsection~\ref{ss:DC-discussion}.}\label{figure2}
\end{figure}

Figure \ref{figure2}, open circles, shows the evolution of the
magnetic induction measured at the centre of the tube, $B_\mathrm{in}
\equiv B_{z\mathrm{in}}(z=0)$, as a function of the applied magnetic
induction. Here, the external source was operated in the DC mode. The
sample described in section \ref{samples} was cooled down to
$77~\mathrm{K}$ in zero-field conditions. Then, we applied an
increasing magnetic induction and reached $B_a=28~\mathrm{mT}$. Upon
decreasing the applied induction to $B_a=-28~\mathrm{mT}$ and
increasing it again up to $B_a=28~\mathrm{mT}$, the internal induction
is seen to follow an hysteretic curve. This behaviour reflects the
dissipation that occurs as vortices sweep in and out of the
superconductor. Remarkably, along the first magnetization curve,
$B_\mathrm{in}$ is negligible below a threshold $B_\mathrm{lim}\approx
14 ~ \mathrm{mT}$ and increases rapidly for higher $B_a$.  As the tube
is no longer an efficient magnetic shield in this latter regime, several
authors regarded $B_{\mathrm{lim}}$ as a parameter determining the
quality of the shield~\cite{110,109,113}. In this paper, we determine $B_\mathrm{lim}$ 
as the maximum applied magnetic induction for which the $DCSF$ is higher than 1000 (60~dB).
In figure \ref{figure2}, $B_\mathrm{lim}$ roughly corresponds to the induction at which the first
magnetization curve meets the hysteretic cycle.

\subsection{Model parameters and numerical results}
\label{ss:DC-discussion}

The shape of the curve of figure \ref{figure2} is indicative of the
 dependence of the critical current density, $J_c$, on the local
 induction. Assuming Kim's model~(\ref{Kim}), the parameters $J_{c0}$ and $B_1$ can be
 extracted from data as follows. First, we neglect flux creep effects
 and set $n\to\infty$. 
 As a result, the current density can either be null or
 be equal to $J=J_c(B)$. Second, we neglect demagnetization effects
 and thus assume that the tube is infinitely long.
 Equation~(\ref{Maxwell-rotB}) then becomes
\begin{eqnarray}
  \frac{\partial B}{\partial r} = \mu_0\,\frac{J_{c0}}{1 + B/B_1}.
\end{eqnarray}
A direct integration yields a homogeneous field in the hollow of
the tube that assumes the form
\begin{eqnarray}\label{Hin}
B_{\mathrm{in}} & = & \left\{
              \begin{array}{cc}
                0 & \textrm{for~}B_a < B_{\mathrm{lim,\infty}},\\
        -B_1+\sqrt{\left(B_1+B_a\right)^2-2 d \mu_0 J_{c0} B_1} &
\textrm{for~}B_a > B_{\mathrm{lim,\infty}},
              \end{array}
             \right.
\label{first-magnetisation}
\end{eqnarray}
where $B_{\mathrm{lim,\infty}}$, defined as
\begin{eqnarray}
B_\mathrm{lim,\infty} = -B_1 + \sqrt{B_1^2 + 2 d \mu_0 J_{c0} B_1}\,,
\label{Bth}
\end{eqnarray}
is the threshold induction assuming an infinite tube with no creep.
Fitting equation (\ref{first-magnetisation}) to experimental data in the region
$B_a > 14~\mathrm{mT}$, we find $B_1=5~\mathrm{mT}$ and
$J_{c0}=1782~\mathrm{A/cm^2}$.

In practice, flux creep effects are present and the exponent $n$
assumes a high, but finite, value. In our case, as to be determined in
the section \ref{ss:AC-mode-simulations}, we found $n\approx 25$. 
The filled squares of figure \ref{figure2} show the simulated values
of the internal induction versus the applied induction, $B_a$, for a tube
with the dimensions of the sample and a flux creep exponent
$n=25$. The $J_c(B)$ relation (\ref{Kim}) was introduced in the equations
of section \ref{ss:algo} with $B_1=5~\mathrm{mT}$ and
$J_{c0}=1782~\mathrm{A/cm^2}$.
These numerical results reproduce the data fairly well. As in the experiment, a
simulated value of $B_\mathrm{lim}$ can be obtained as the maximum applied induction
for which the DCSF is higher than 60~dB. We also obtain $B_{\mathrm{lim}}\approx 14~\mathrm{mT}$.
We note that even in the presence of flux creep with $n=25$, the
simulated $B_\mathrm{lim}$ has the same value as the one
given in Kim's model, (\ref{Bth}).

\subsection{Modelling of the field penetration into a HTS tube}
\label{ss:DC-flux-distribution}

In this section, we compare the penetration of the magnetic flux
in a tube and in a bulk cylinder through a numerical analysis. This comparison reveals the
coexistence of different penetration mechanisms in the tube. An
understanding of these mechanisms is necessary to predict the efficiency
of a HTS magnetic shield.

We use the numerical model introduced in section \ref{compute}, with a flux creep
exponent $n=25$. In order to facilitate comparisons with results from
the literature, we choose the critical current
density, $J_c$, to be independent of the local magnetic induction. 
We further wish to normalize the applied field to the full penetration
field, $H_P$, that, in the limit $n\rightarrow\infty$, corresponds to the
field for which the sample is fully penetrated and a current density $J_c$ flows throughout the
entire volume of the superconductor.

For a bulk cylinder of radius $a_2$ and length $\ell$, $H_P$ assumes the form~\cite{169}:
\begin{equation}\label{Hpb}
H_P={J_c \ell\over 2}\:\ln
\left(\frac{2 a_2}{\ell}+\sqrt{1+\frac{4 a_2^2}{\ell^2}}\right). %eq 11 de [169]
\end{equation}
In the limit $\ell\rightarrow\infty$, one recovers the Bean limit 
$H_{P\infty} = J_c a_2$.
An approximate expression of $H_P$ for a tube can
be obtained with the energy minimization approach developed in \cite{164}:
\begin{equation}\label{Hp tube OK}
H_P=J_c {\ell\over 2}\:\frac{1-\delta}{1+\delta}\:\ln\left[\frac{2
a_2(1+\delta)}{\ell}+
\left(1+\left(\frac{2 a_2 (1+\delta)}{\ell}\right)^2\right)^{1/2}\right],
\end{equation}
with $\delta=a_1/a_2$. 
An interesting observation is that (\ref{Hp tube OK}) can be rewritten as:
\begin{equation}\label{Hptube}
H_P=J_c d \frac{\ell}{4\bar{a}}\:\ln\left[\frac{4
\bar{a}}{\ell}+
\left(1+\left(\frac{4 \bar{a} }{\ell}\right)^2\right)^{1/2}\right]
\end{equation}
where $\bar{a}=(a_1+a_2)/2$ is the mean radius. This shows that
the correction to the field $H_P$ of an infinite tube, $H_{P\infty} = J_c (a_2 - a_1)=J_c d$,
depends only on the ratio $\ell/\bar{a}$. Physically, this ratio is a measure of the importance
of end effects.

\begin{figure}[ht!]
\center \includegraphics[scale=0.8]{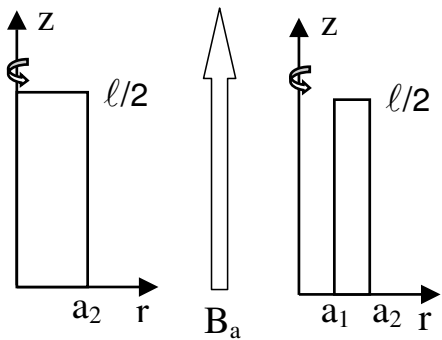}\caption {Cylinder
and tube of external radius $a_2$, and length $\ell =6~a_2$ subjected to
an axial magnetic induction $\mathbf{B}_a=B_a\hat{z}$. Only the region
$0\leq r \leq{a_2}$ and $0\leq{z}\leq{\ell/2}$ is depicted for symmetry reasons.}\label{figure3}
\end{figure}

Consider then the cylinder and the tube
of figure~\ref{figure3}, both of external radius $a_2$ and length
$\ell = 6a_2$. The inner radius of the tube is $a_1=0.5a_2$.  Both
samples are subjected to an increasing axial magnetic induction, with
$\dot{B}_a(t)=E_c/a_2$ and $B_a(0)=0$.

\begin{figure}[h!]
\center \includegraphics[scale=0.5]{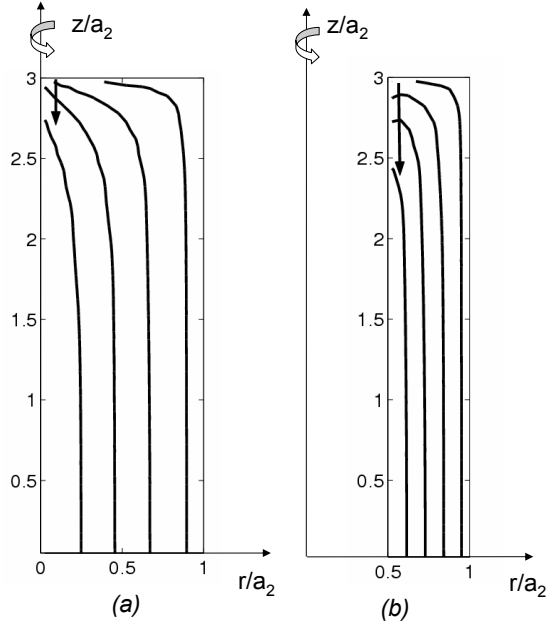}\caption
{Cylinder (a) and tube (b) of external radius $a _2$, of length
$\ell=6~a_2$ subjected to an axial magnetic field. The internal radius of
the tube is $a_1=0.5~a_2$. The samples are characterized by field independent
 $J_c$ and $n$ values ($n=25$). As flux
lines are symmetric about $z=0$ and $r=0$, only the region $0\leq{r}\leq{a_2}$
and $0\leq{z}\leq {\ell/2}$ is depicted. The contour curves show the flux
fronts at $B_a=0.1,0.3,0.5,0.7~\mu_0 H_{P}$ where $H_{P}$ is the field of
full penetration.}\label{figure4a_4b}
\end{figure}

Figure \ref{figure4a_4b} shows a comparison of the simulated flux
front for the cylinder (a) and for the tube (b) as a function of the
applied magnetic induction. Here, the flux front corresponds to the locus of positions at which the current density rises to $J_c/2$. 
To label the front as a function of the applied induction, we have taken as a reference magnetic 
field the full penetration field, $H_{P}$, whose expression is given in (\ref{Hpb}) and (\ref{Hp tube OK}), both for the bulk cylinder and for the tube.
The flux front is depicted for different external magnetic inductions with
$B_a/(\mu_0 H_P) = 0.1, 0.3, 0.5,~\textrm{and}~0.7$. We note that the front
shapes are similar to those obtained by Navau et al.~\cite{164}, which
used an approximate method based on the minimization of the total
magnetic energy to study the field penetration into bulk and hollow cylinders.
Due to the finite length of the samples, the flux
fronts are curved in the end region $z\cong \ell/2$. Remarkably, this
curvature implies that the magnetic flux progresses faster towards
$z=0$ along the inner boundary of the tube ($r=a_1$) than the magnetic
flux penetrates the central region near $z=0$ in a bulk
cylinder. Thus, two penetration mechanisms coexist for the tube: the
magnetic field can penetrate either from the external boundary at
$r=a_2$, as in the cylinder, or from the internal boundary at $r=a_1$,
via the two openings.

\begin{figure}[ht!]
\center \includegraphics[scale=0.5]{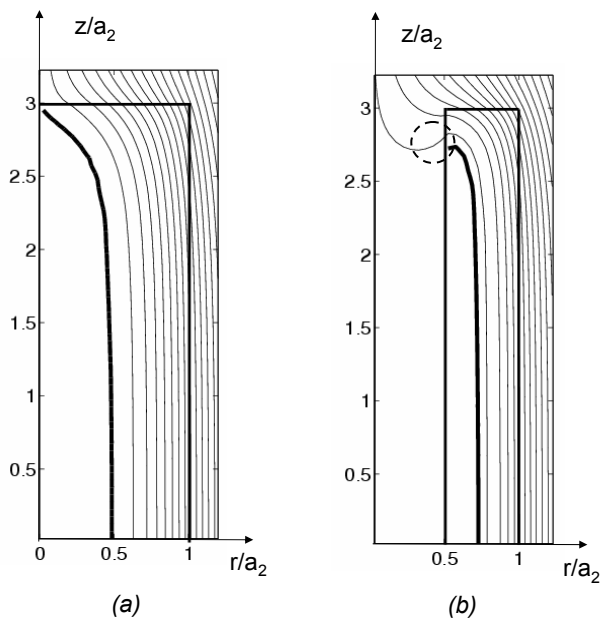}\caption {Cylinder
(a) and tube (b) of external radius $a_2$ and length $\ell=6a_2$
subjected to an axial magnetic field. The samples are characterized by
$J_c$ and $n$ values independent of the local magnetic induction ($n=25$). Only the region $0\leq{r}\leq{a_2}$ and
$0\leq{z}\leq{\ell/2}$ is depicted. The applied induction is
$0.5~\mu_0 H_P$. The thick line represents the flux front ($J=J_c/2$) and
the thin lines represent the magnetic field lines. One can observe negative components $B_z$ in the dashed circle of figure (b).}\label{figure5a_5b}
\end{figure}

Consider next the field lines~\footnote {A general difficulty arises when one tries to visualize
3D magnetic field lines with axial symmetry in a 2D plot. Here, we
have used contours of the vector potential $A(r,z)$ at equidistant
levels. Another possibility would be to use contours of $rA(r,z)$ at
non-equidistant levels. Brandt has shown~\cite{144} that both
approaches provide reasonably good approximations of the field
lines.}
for the cylinder and for
the tube submitted to axial fields equal to half of their respective
field $H_{P}$ (see figure \ref{figure5a_5b}). The shape of the field lines in
the region near $z=b$ are seen to be very different for the cylinder
and for the tube. In particular, for the tube, the component $B_z$ is
negative near the opening and close to the inner boundary, as seen in the dashed circle of figure \ref{figure5a_5b}(b).  Such a
behaviour is reminiscent of the field distribution found in the
proximity of a thin ring~\cite{156, 158, 157, 155}.

The existence of a negative $B_z$ inside the hollow part of the tube can be
interpreted as follows. For an infinitely long tube, the magnetic
field can only penetrate from the external surface and the field lines
are parallel to the axis of the tube. As the length $\ell$ of the tube
decreases, the flux lines spread out near $z=\ell$ due to demagnetization
effects. As a result, shielding currents in the end region of the
tube fail to totally shield the applied field and a non-zero magnetic
field is admitted through the opening. The shielding currents flow in
an extended region in the periphery of the superconductor. In the
superconductor, ahead of the flux front, there is no shielding
current and hence no electric field. Integrating Faraday-Lenz's law
along a contour lying in a non-penetrated region thus gives zero,
meaning that the flux threaded by this contour must also be null. (As
a reminder, the sample is cooled in zero-field.) Therefore, the
magnetic flux due to the negative component $B_z$ near $r=a_1$ is
there to cancel the positive flux that has been allowed in the hollow
of the tube near the axis.

%-----------------------------------------------------
\subsection{Uniformity of the field attenuation in a 
superconducting tube}\label{uniformity}
%-----------------------------------------------------

Since magnetic flux can penetrate both through the outside surface and through the
openings, it is therefore relevant to investigate how the magnetic
induction varies in the hollow of the tube. 
Numerical simulations show that the variation of the field attenuation along the radius is much smaller than the variation along the $z$-axis. We thus concentrate on the latter and study the DC shielding factor, $DCSF$, as a function of $z$. 

\begin{figure}[ht!]
\center \includegraphics[scale=0.75]{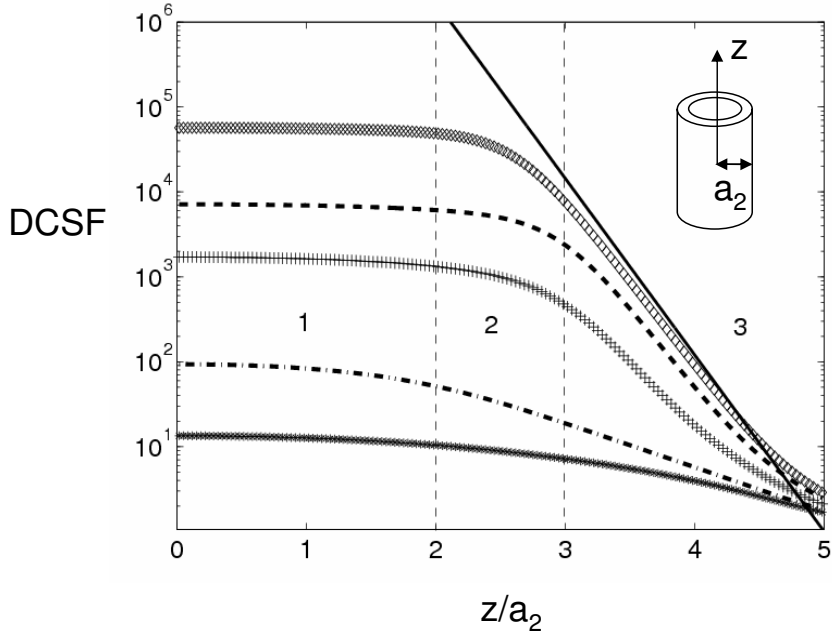}\caption {Simulated
variation of the DC shielding factor along the $z$-axis for increasing applied
inductions. From top to bottom : $B_a/B_{\mathrm{lim}}
=0.8,0.9,1,1.05,\:\texttt{and}\: 1.1$. The threshold $B_\mathrm{lim}$ is
determined as the maximum applied induction for which the shielding factor is higher than 60~dB.  The geometry is identical to that of the
sample studied experimentally. The black solid line is the equation $DCSF(z)=
e^{C(\ell/2-z)/a_1}$ (see text). For symmetry reasons, only the upper half of the
tube is shown.}
\label{figure6}
\end{figure}

Figure \ref{figure6} shows the variation of
$DCSF$ along the $z$-axis as a
function of the external induction $B_a$. 
The geometrical parameters are those of the sample studied experimentally
and a $J_c(B)$ relation with the parameters of section \ref{ss:DC-discussion} is used. 
As the curve $DCSF(z)$ is
symmetric about $z=0$, only the portion $z>0$ is shown. Three
different behaviours can be observed: in region $1$, the shielding
factor is nearly constant; in region $2$, it starts decreasing
smoothly; it falls off as an exponential in region $3$, which is
roughly defined as the region for which $z > \ell/2 - 2 a_2$.

A useful result is known for semi-infinite tubes made of type-I
superconductor and subjected to a weak axial field. In the Meissner
state, the magnitude of
the internal induction, $B_{\mathrm{in}}$, decreases from the extremity of the
tube~\cite{177} as
\begin{equation}\label{Cabrera}
B_{\mathrm{in}}\propto  \mathrm{e}^{-C (\ell/2-z)/a_1}, %eq 1.44 de [177]
\end{equation}
where $a_1$ is the inner radius, and $C \approx 3.83$ is the first zero of the Bessel
function of the first kind $J_1(x)$. This result holds for $\ell/2-z\gg
a_1$ and implies that the shielding factor increases as an exponential
of $\ell/2-z$.  An exponential dependence has also been measured in some
HTS materials 
for applied fields above $H_{c1}$~\cite{173,112}. Other
measurements~\cite{196} in similar conditions have shown instead a
uniform shielding factor in a region around the center of the tube.

From the simulation results 
we see that both behaviours can actually be observed in a type-II
tube, provided the ratio $\ell/\bar{a}$ is large. For the sample
studied in this paper, this ratio is equal to
$\ell/\bar{a}\sim 11$.  The
exponential falloff approximately follows the law $DCSF(z)\sim
\exp(C(\ell/2 - z)/a_1)$ (black solid line) for the lowest fields only,
but appears much softer for the larger magnitudes $B_a$. This behaviour
can be attributed to the fact that as $B_a$ increases, the region near
$z=\ell/2$ becomes totally penetrated (see figure~\ref{figure4a_4b}) and
the ``effective'' length of the tube decreases. It leads in turn to a
reduction of the distance to the extremity, $\ell/2 - z$, which therefore
softens the falloff of the shielding factor.

%144e/tube12/SFybis

\begin{figure}[ht!]
\center \includegraphics[scale=0.75]{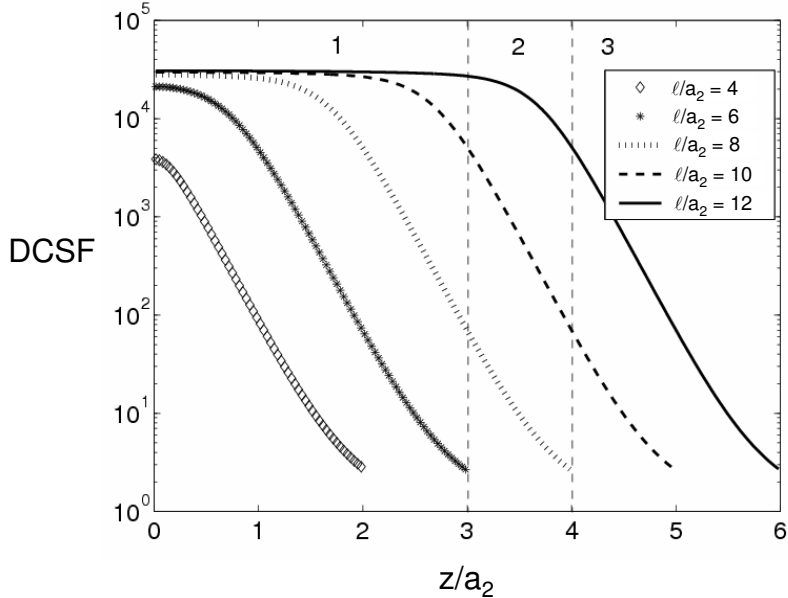}\caption {Evolution of the
DC shielding factor along the axis for different lengths,
$\ell$. The internal and external radii, $a_1$ and $a_2$,
are kept fixed ($a_1=0.8~a_2$). The applied induction is
$B_a=0.85~B_{\mathrm{lim}}$.}~\label{figure7}
\end{figure}

The two behaviours --- a nearly constant shielding factor and
an exponential decrease of this factor --- can be associated with the
two mentioned penetration mechanisms. For the part of the flux that penetrates
via the openings, we expect the shielding factor to increase as an
exponential of $(\ell/2-z)$ as one moves away from the
extremity. This is the behaviour observed in type-I shields, for which
no flux can sweep through the side wall if $d \gg \lambda$, where $\lambda$ denotes the 
London penetration depth. By contrast, in the centre region, for a tube with a large $\ell/\bar{a}$
ratio, the flux penetrating via the openings is
vanishingly small and flux penetration through the walls
prevails. This leads to the nearly constant shielding factor observed
in region $1$. As the ratio $\ell/\bar{a}$ increases, flux penetration
through the wall strengthens. As a result, the plateau region increases
in size as is confirmed in figure~\ref{figure7}, which shows $DCSF(z)$
for six different lengths $\ell/a_2$ (the outer radius and the width $d=a_2-a_1=0.2a_2$ are
kept fixed) and for $B_a = 0.85~B_{\mathrm{lim}}$. 
Note that the plateau of the shielding factor
disappears for the smallest ratios $\ell/\bar{a}$ (for
$\ell\leq 6a_2$) as for these ratios, flux penetration through the
openings competes with that through the wall. 

This last example shows that it is important to distinguish $B_{\mathrm{lim}}$, which we have defined as the maximum induction for which $DCSF$ is larger than 60~dB, from $\mu_0 H_P$, which corresponds to the full penetration of the sample. 
In fact, for $\ell<6a_2$, the attenuation falls below 60~dB before the sample is full penetrated.
If $\ell$ is further reduced, $\ell\leq2a_2$, it is actually not possible to define an induction
$B_{\mathrm{lim}}$, as $DCSF$ is lower than 60~dB for any applied inductions. 
Therefore, the interest of using short open HTS tubes for magnetic shielding applications
is very limited.

%144i/tube20/SFz.fig : p18prime de 144i

When $\ell\geq6a_2$, the value of $B_{\mathrm{lim}}$ 
is very close to the applied field for which the sample is fully penetrated, as the main 
penetration mechanism is the non-linear diffusion through the superconducting wall. 
To evaluate $B_{\mathrm{lim}}$, one could then use (\ref{Hp tube OK}), which for $\ell\geq 6a_2$, is close to $H_{P\infty}=J_cd$ . However, this formula
can be misleading for understanding the influence of the wall thickness, $d$. Expressions 
(\ref{Hp tube OK}) or $H_{P\infty}=J_cd$ were established ignoring the variation of $J_c$ with $B$ and show a linear 
dependence of $B_{\mathrm{lim}}$ as a function of $d$. However, the decrease of $J_c$ with the local induction
yields a softer dependence as can be seen in (\ref{Bth}).
There, $B_{\mathrm{lim}}\approx B_{\mathrm{lim},\infty}$ is linear in $d$ only for
thicknesses $d$ much smaller than $B_1/(2 \mu_0 J_{c0}) \approx
0.1~\mathrm{mm}$, but grows as $\sqrt{d}$ for larger
thicknesses if one takes the $J_{c0}$ and $B_1$ parameters of section \ref{ss:DC-discussion}.
Thus, if one wants to shield high magnetic inductions
(larger than $100~\mathrm{mT}$) with a superconductor similar to that
described in section \ref{samples}, unreasonnably thick wall
thicknesses are required.  In this case, it is advisable to first
reduce the field applied to the superconductor by placing a
ferromagnetic screen around it.

A final remark concerns the effect of the width of the superconducting wall, $d$, on the spatial
dependence of $DCSF$. If $d$ is increased while the ratio
$B_a/B_\mathrm{lim}$ is kept fixed, the shielding factor increases in
magnitude but its $z$-dependence remains qualitatively the same.

In this section, we used a quasistatic field. The results concerning the spatial variation of the field attenuation are expected to be still valid in the case of an AC field.

\section{Magnetic shielding in the AC mode}
\label{s:AC-mode}

The sensing coil of the setup described in section \ref{samples} can move along the axis of the sample. In this section, we first present the measured variation of the AC shielding factor along the axis of the tube and compare it to numerical simulations for which an AC applied induction is used.
We also measure the frequency
response and interpret the results with scaling laws arising from the constitutive law $E\propto J^n$.

\subsection{Experimental results}
\label{ss:AC-mode-experiment}

\begin{figure}[ht!]
\center \includegraphics[scale=1.2]{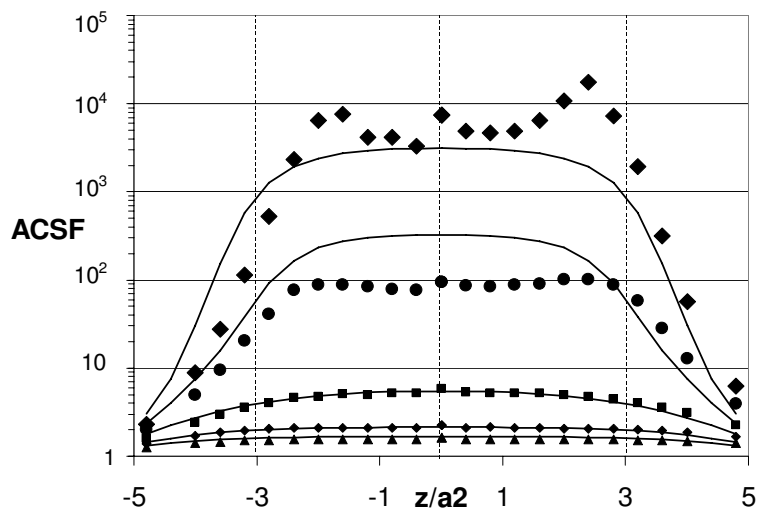}\caption {Variation
of the AC shielding factor along the axis of the sample, at a
frequency $f=103~\mathrm{Hz}$. Filled symbols~:~measurement. Continuous lines~:~simulation. From top to bottom: $B_{a,RMS}= 10.8,\,12,\,13.4,\,15.3,\,\texttt{and}\,16.6~\mathrm{mT}$. }
\label{figure8}
\end{figure}

The variations of the measured AC shielding factor $ACSF$ defined in (\ref{SFAC}), along the axis of the sample studied experimentally 
for a fixed frequency and varying
amplitudes of the applied field are shown in figure~\ref{figure8} (filled symbols).
Apart from the upper curve of figure \ref{figure8} corresponding to 
$B_{a,RMS}=10.8~\mathrm{mT}$, we observe a
nearly constant measured shielding factor in the central region . Going further to the
extremity of the tube, near $z = 5a_2$, $ACSF$ decreases as an
exponential.

\begin{figure}[ht!]
\center \includegraphics[scale=1.2]{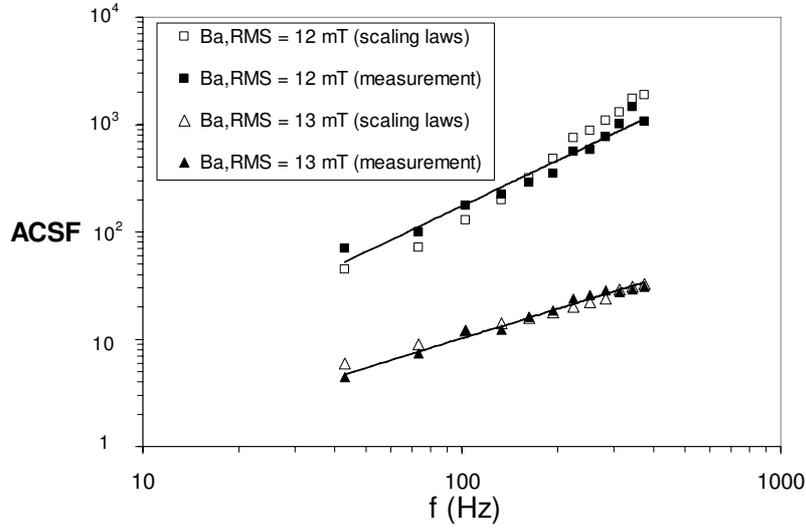}\caption {AC shielding
factor versus frequency. The filled symbols come from a direct
measurement and the open symbols correspond to an estimation based on
scaling laws (see text). The two lines show that the variation of the AC shielding
factor with the frequency is close to a power law.}
\label{figure9}
\end{figure}

Figure \ref{figure9} (filled symbols) shows a measurement of the AC shielding factor, $ACSF$, as a function of
frequency for two applied magnetic inductions when the magnetic sensor is
placed at the centre of the sample.
The frequency dependence appears to follow a power law. 

\begin{figure}[ht!]
\center \includegraphics[scale=1.0]{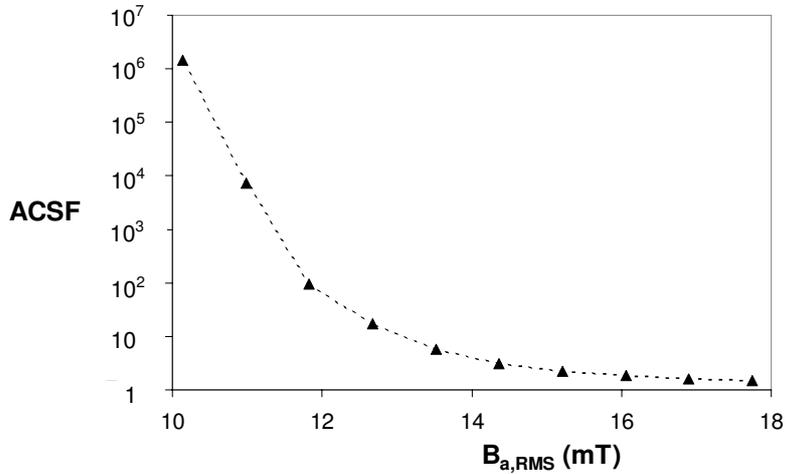}\caption {Measured AC
shielding factor at the centre of the sample versus the RMS value of the
applied magnetic induction $B_{a,RMS}$. Its frequency is kept fixed at
$f=103~\mathrm{Hz}$.} \label{figure10}
\end{figure}

Figure
\ref{figure10} shows the evolution of the AC shielding factor 
measured at the centre of the tube at a fixed
frequency $f=103~\mathrm{Hz}$ and for varying RMS values of the applied induction.
The shielding factor decreases with $B_{a,RMS}$.

%/mesures/blind11/bas.xls

\subsection{Uniformity of the field attenuation}
\label{ss:AC-mode-space}
The solid lines of figure~\ref{figure8} represent the simulated $ACSF$
for the applied inductions used during the measurement. As in the DC case, 
we observe a constant shielding factor around the centre $z=0$ of the tube whereas
$ACSF$ falls off exponentially near the opening ends $z=5a_2$. Remarkably, one
can observe the relative good quantitative agreement between simulated and 
experimental results of figure~\ref{figure8}. 
For $B_{a,RMS}=10.8~\mathrm{mT}$, local variations of the measured $ACSF$ can be observed
for $\left|z\right|<3$. In particular, the maximum shielding factor is
no longer located at the centre of the tube, and shielding appears to
be asymmetric in $z$. For higher values of the applied magnetic induction, the
maximum $ACSF$ lies at $z=0$ and shielding recovers its symmetry
about the centre. These effects are supposed to be due to non-uniform superconducting properties.

\subsection{Scaling laws and frequency response}
\label{ss:AC-mode-simulations}

The strong non-linearity of the constitutive law $E = E_c (J/J_c)^n$
gives rise to frequency scaling laws with $n-$dependent power
exponents~\cite{144}. The scaling laws can be obtained by changing the
time unit in Maxwell's
equations (\ref{Maxwell-rotE}) and (\ref{Maxwell-rotB}) by a factor $c>0$, $t \mapsto t_{\mathrm{new}} = {t / c}$. Given a solution with a current density $J(\mathbf{r},t)$, an applied
induction $B_a(\mathbf{r},t)$, and a total induction
$B(\mathbf{r},t)$, new solutions can be found that satisfy
\begin{eqnarray}
J_{\mathrm{new}}(\mathbf{r},t_{\mathrm{new}})&=&J(\mathbf{r},t)\,c^{1/(n-1)}, \label {Jtild}\\
B_{\mathrm{new}}(\mathbf{r},t_{\mathrm{new}})&=&B(\mathbf{r},t)c\,^{1/(n-1)}, \label {Btild}\\
B_{a,\mathrm{new}}(\mathbf{r},t_{\mathrm{new}})&=&B_a(\mathbf{r},t)\,c^{1/(n-1)}.\label {Batild}
\end{eqnarray}
Transposed to the frequency domain, these relations imply that if the
frequency of the applied field is multiplied by a factor $c$, then the
current density and the magnetic induction are rescaled by the factor
$c^{1/(n-1)}$. In particular, if one knows the $ACSF$ corresponding to the applied induction
$B_a$ at the frequency $f$, $ACSF(Ba,f)$, one can deduce the $ACSF$ corresponding to
the magnetic induction  $B_{a,\mathrm{new}}$ at the frequency $f_{\mathrm{new}}=c\,f$ by using:
\begin{equation}\label{ACSFnew}
ACSF(B_{a,\mathrm{new}},f_{\mathrm{new}})=ACSF(B_a,f),
\end{equation} 
as $ACSF$ is the ratio of two magnetic inductions (see (\ref{SFAC})) and is thus invariant under scaling.
Then, the frequency dependence of $ACSF$ in figure \ref{figure9} can be reproduced as follows using these scaling laws. 
First, we approximate the curve of figure \ref{figure10} by piecewise exponentials, which gives
$ACSF(B_{a},103~\mathrm{Hz})$. 
Second, we use (\ref{ACSFnew}) and write:
\begin{equation}\label{scaling}
ACSF(12~\mathrm{mT},f_{\mathrm{new}})=ACSF(B_a,103~\mathrm{Hz}),    
\end{equation}
with
\begin{eqnarray}\label{Ba}
B_a &= & B_{a,\mathrm{new}}\,c^{-1/(n-1)}\\
 &=&    12~\mathrm{mT}\left(\frac{103}{f_{\mathrm{new}}}\right)^{1/(n-1)}
\end{eqnarray}
Hence, the variations with respect to $B_a$ in figure \ref{figure10} can
be translated into frequency variations at a fixed induction. 
This gives the upper curve of figure \ref{figure9} (open symbols) for which
we used $n=25$. The lower curve
is obtained by fixing $B_{a,\mathrm{new}}$ to $13~\mathrm{mT}$.
This construction thus demonstrates that the frequency
variation intrinsically arises from scaling laws.

The detailed construction relies on a specific value of the creep
exponent $n$, which we have taken here to be equal to $n=25$ and
independent on $B$. Analysing the frequency dependence with
scaling laws thus also serves the purpose of determining the value of
$n$ that best fits experimental data.
A HTS shield characterized by a lower $n$ value would present a more pronounced 
frequency dependence of the shielding factor.

One may wonder on the role played by the increased dissipation, due to
the motion of vortices, as frequency is increased. Such dissipation
can lead to a temperature rise, a decrease of the critical current
density, and thus a decrease of the shielding factor.  Nevertheless,
it appears from figure \ref{figure9} that the temperature increase must
remain small in the frequency window investigated 
in our experiment ($43~\mathrm{Hz}-373~\mathrm{Hz}$), as no significant
reduction of $ACSF$ can be observed in that frequency range.
One may equally wonder on the role played by the different harmonics of the internal magnetic induction. For the applied fields we consider here, the fundamental component strongly dominates the higher harmonics. As as consequence, the curves of figures \ref{figure9}, and \ref{figure10} are not significantly affected if one takes the RMS value of the total internal magnetic induction, rather than its fundamental component, to define the shielding factor in the AC mode.

%-----------------------------------------------------
\section{Conclusions} \label{conclusions}
%-------------------------------------------------
We have presented a detailed study of the magnetic shielding properties of a 
polycrystalline Bi-2223 superconducting tube subjected to an axial field. We have measured the field attenuation with high sensitivity for DC and AC source fields, and have confronted data with computer modelling of the field distribution in the hollow of the tube. The numerical model is based on  the algorithm described in \cite{144}, which is easy to implement on a personal computer. Our study allows us to detail the variation of the shielding factor along the axis, interpret it in terms of the penetration mechanisms, and take into account flux creep and its effect on the frequency dependence. To our knowledge, it is the first study which systematically describes the spatial and frequency variations of the shielding factor in the hollow of a HTS tube. 

Our main findings can be summarized as follows. 
\begin{itemize}
\item A HTS tube can efficiently shield an axial induction below a threshold induction $B_\mathrm{lim}$. For our commercial sample, $B_\mathrm{lim}= 14~\mathrm{mT}$. The threshold induction $B_\mathrm{lim}$ increases with the ratio $\ell/\bar{a}$, the thickness of the tube, and depends on the exact $J_c(B)$ dependence ($\ell$ is the length of the tube and $\bar{a}$ is the mean radius). When the length of the tube decreases, $B_\mathrm{lim}$ can be strongly reduced because of demagnetizing effects. 
\item There are two penetration mechanisms in a HTS tube in the parallel geometry: one from the external surface of the tube, and one from the opening ends, the latter mechanism being suppressed for long tubes. 
These two mechanisms lead to a spatial variation of the shielding factor along the axis of the tube. 
In a zone extending between $z=0$ (centre of the tube) and $z=\ell/2-3a_2$, the shielding factor is constant when $\ell>6a_2$ ($a_2$ is its external radius). Then it decreases as an
exponential as one moves towards the extremity of the tube. As a consequence of this spatial dependence, no zone with a constant shielding factor exists for small tubes ($\ell <6 a_2$).  
\item The shielding factor increases with the frequency of the field to shield, following a power law. This dependence can be explained from scaling laws arising from the constitutive law $E\propto J^n$.
\end{itemize}

In practice, a tube of a Bi-2223 ceramic can thus be used to effectively shield an axial field at low frequencies. A sample with an outer radius $a_2=1.8~\mathrm{mm}$, a length $\ell > 6 a_2$, a thickness $d=1.5~\mathrm{mm}$, and with superconducting properties similar to the ones of our sample (table~\ref{t:sample-description}),  strongly attenuates magnetic inductions lower than $B_\mathrm{lim} = 14~\mathrm{mT}$ at $77~\mathrm{K}$. The shielding factor is nearly constant and larger than $10^3$ (60~dB) in the region $|z| < \ell/2-3a_2$ if the applied induction is lower than $0.9~B_\mathrm{lim}$.

%----------------------
\section{Acknowlegment}
E.H.~Brandt is gratefully acknowledged for useful discussions. A.F.~Gerday and D.~Bajusz are also acknowledged for their experimental support. This research was supported in part by a
ULg grant (Conseil de la Recherche support through project ``Fonds sp{\'e}ciaux'' (C-06/03)) and by the Belgian F.N.R.S (grant from FRFC: 1.5.115.03).

%--------------------------------------------------------------------
% Référence, Bibliographie
%--------------------------------------------------------------------

{}

%--------------------------------------------------------------------
% Index
%--------------------------------------------------------------------

\end{document}